\documentstyle[12pt]{article}

\begin{document}

\begin{flushright}
IMSc/2007/07/09
\end{flushright} 

\vspace{2mm}

\vspace{2ex}

\begin{center}

{\large \bf Consequences of U dualities for } \\ 

\vspace{2ex}

{\large \bf  Intersecting Branes in the Universe  } \\

\vspace{8ex}

{\large  S. Kalyana Rama}

\vspace{3ex}

Institute of Mathematical Sciences, C. I. T. Campus, 

Tharamani, CHENNAI 600 113, India. 

\vspace{1ex}

email: krama@imsc.res.in \\ 

\end{center}

\vspace{6ex}

\centerline{ABSTRACT}
\begin{quote} 

We consider N -- charge, intersecting brane antibrane
configurations in M theory which are smeared uniformly in the
common transverse space and may describe our universe. We study
the consequences of U dualities and find that they imply
relations among the scale factors. We find using Einstein's
equations that U dualities also imply a relation among the
density $\rho$ and the pressure $p_i$ for the single charge
case. We present an ansatz for $\rho$ and $p_i$ for the N --
charge case which yields all the U duality relations among the
scale factors. We then study configurations with identical
charges, and also with net charges vanishing. We find among
other things that, independent of the details of the brane
antibrane dynamics, such four charge configurations lead
asymptotically to an effective $(3 + 1)$ -- dimensional
expanding universe.

\end{quote}

\vspace{2ex}


\newpage

\vspace{4ex}

{\bf 1. Introduction}  

\vspace{2ex}

It is important to understand how our $(3 + 1)$ -- dimensional
universe may be described within string theory or, equivalently,
within M theory. Enormous amount of work has been done
addressing this issue, a sample of which is given in
\cite{bowick} -- \cite{k2}.

Chowdhury and Mathur proposed recently that mutually BPS, multi
charge, intersecting brane antibrane configurations in M theory,
smeared uniformly in the common transverse space, may describe
our universe \cite{cs}. This is because the branes, and
similarly antibranes, in such configurations form bound states,
become fractional, support very low energy excitations, and thus
have high entropy. In this paper, therefore, we consider such N
-- charge configurations with three or more common transverse
directions.

We take the brane directions to be toroidal and study the
consequences of U dualities for such configurations. U dualities
here refer to suitable combinations of dimensional reduction,
dimensional uplifting, S and T dualities. We find that U
dualities imply relations among the scale factors, which then
are characterised by N independent functions.

The energy momentum tensor, $T^\mu \; _\nu = diag \; (- \rho, \;
p_i) \;$, for such a configuration may be determined, in
principle, by brane antibrane dynamics. We find using Einstein's
equations that, as a consequnce of U duality, $\rho$ and $p_i$
for the single charge case obey a relation. We then present an
ansatz for $\rho$ and $p_i$ for the N -- charge case which
yields all the U duality relations among the scale factors.

Using this ansatz, we study configurations with identical
charges and find, among other things, that four charge
configurations lead asymptotically to an effective $(3 + 1)$ --
dimensional expanding universe. This result follows as a
consequence of U dualities in M theory and is independent of the
details of the brane antibrane dynamics.

We also study configurations with identical charges and with net
charges all vanishing which, for entropic reasons, are likely to
dominate the universe. Assuming a particular equation of state,
we list the asymptotic solutions for a few such N -- charge
configurations and discuss some of their properties. We find
that four charge configurations are likely to dominate the
universe and may provide a detailed realisation of the maximum
entropic principle that we have proposed recently in \cite{k2}
to determine the number $(3 + 1)$ of large spacetime
dimensions. Even otherwise, these configurations provide, at the
least, a model for our $(3 + 1)$ -- dimensional expanding
universe.

This paper is organised as follows. We discuss the consequences
of U duality for scale factors in section {\bf 2}, and for
energy momentum tensor in section {\bf 3}. We then study
configurations with identical charges in section {\bf 4}, and
those with net charges also vanishing in section {\bf 5}. We
present a brief summary and then conclude by mentioning a few
issues for further study in section {\bf 6}.

\vspace{4ex}

{\bf 2. Consequences of U dualities for scale factors}  

\vspace{2ex}

Mutually BPS, non extremal, intersecting branes in M theory can
be thought of as intersecting brane antibrane configurations
\cite{hm}, and describe black holes when localised in the common
transverse space. A configuration with $N$ -- types of branes
\footnote{ Here and in the following, we use the terms `branes',
`brane antibrane', `branes and antibranes', et cetera
interchangeably and also use them to mean `waves', `wave
antiwave', `waves and antiwaves', et cetera. A wave antiwave
configuration is that obtained, for example, from M2 brane
antibrane configuration by an appropriate U duality. The
configurations considered in this paper always consist of branes
and antibranes and/or waves and antiwaves.  Hence, our
interchangeable use of the terms above is unlikely to cause any
confusion; also their intended meaning will be clear from the
context.} will be referred to as $N$ -- charge configurations.
These branes intersect as per the rules given in \cite{bps,
addn}, form bound states, become fractional, support very low
energy excitations, and thus have high entropy. Chowdhury and
Mathur proposed recently that such intersecting configurations,
when smeared uniformly in the common transverse space, may
describe our universe \cite{cs}. In the following, we consider
only such configurations with three or more common transverse
directions.

$N$ -- charge intersecting configurations in M theory can be
transformed into each other by suitable combinations of
dimensional reduction, dimensional uplifting, S and T dualities,
collectively referred to in the following as U dualities. Let
$\downarrow_i$ and $\uparrow_i$ denote dimensional reduction and
uplifting along $i^{th}$ direction between M theory and type IIA
string theory; $T_j$ denote T duality along $j^{th}$ direction
in type IIA/B string theories; and $S$ denote S duality in type
IIB string theory. Then the U duality $\uparrow_i T_j S T_j
\downarrow_i \;$ interchanges $i$ and $j \;$. The U dualities of
the type $\uparrow_i T_j T_k \downarrow_i \;$ transform one $N$
-- charge configuration to another.

For example, the U duality $\uparrow_3 T_4 T_5 \downarrow_3 \;$
transforms the $N = 1$ configuration $2 : 12$ to $5 : 12345 \;$,
whereas $\uparrow_3 T_1 T_2 \downarrow_3 \;$ transforms it to $W
: 3 \;$; and, $\uparrow_5 T_1 T_2 \downarrow_5 \;$ transforms
the $N = 4$ configuration $2255 : (12, 34, 13567, 24567)$ to
$W555 : (5, 12345, 23567, 14567) \;$. \footnote{ Our notation
for the brane configurations is as follows: In configuration
2255, there are two types of M2 branes along the directions
$(x^1, x^2)$ and $(x^3, x^4) \;$, and two types of M5 branes
along $(x^1, x^3, x^5, x^6, x^7)$ and $(x^2, x^4, x^5, x^6,
x^7)$; and similarly for the corresponding antibranes. In W555,
there are three types of M5 branes along the directions
indicated and also a wave along $x^5 \;$. Similarly for other
configurations.}

In this paper, we assume that $N$ -- charge intersecting brane
antibrane configurations are smeared uniformly in the common
transverse space and describe our universe. The spatial
directions parallel to the branes are taken to be toroidal. The
common transverse directions may also be taken to be toroidal,
and with sufficiently large radii so as to describe our
universe, or may simply be taken to be non compact. The
corresponding line element $d s$ is given by
\begin{equation}\label{ds} 
d s^2 = - e^{2 \lambda_0} d \tilde{t} \; ^2 
+ \sum_i e^{2 \lambda_i} d x_i^2
\end{equation} 
where $i = 1, 2, \cdots, 10$ and $\lambda_i$ depend only on
time. \footnote{If there is a wave along $x$ then $d x$ in
equation (\ref{ds}) stands for $d x - A_\mu d x^\mu$ where
$A_\mu$ is to be obtained from $A_{\mu \nu \rho}$ by the
corresponding U duality. Such gauge fields will be incorporated
in the energy momentum tensor, which will be given later by an
ansatz. Explicit expressions for gauge fields are not needed
then and, hence, will not be shown here although it is an
interesting problem to obtain the gauge fields corresponding to
the given ansatz.} Note that the physical time $t$ is given by
$d t = e^{\lambda_0} d \tilde{t}$ and that one may set
$\lambda_0 = 0$ with no loss of generality. Hence, in the
following, we will not keep track of $\lambda_0$ and its
transformations under U dualities. 

We now study the implications of U dualities of the type
$\uparrow_i T_j T_k \downarrow_i \;$ which transform $N$ --
charge configurations into each other. Under such a U duality,
it can be shown that $\lambda_i$ in equation (\ref{ds})
transform to $\lambda_i'$ given by
\begin{eqnarray}
& & \lambda_i' = \lambda_i - 2 \lambda \; \; , \; \; \; 
\lambda_j' = \lambda_k - 2 \lambda \; \; , \; \; \; 
\lambda_k' = \lambda_j - 2 \lambda \nonumber \\
& & \lambda_l' = \lambda_l + \lambda \; \; , \; \; \;
l \ne i, j , k \; \; \; ; \; \; \;
\lambda \equiv \frac{\lambda_i + \lambda_j + \lambda_k}{3} 
\; \; .  \label{ijk}
\end{eqnarray}
Consider, as an example, the configurations $2 : 12$ with scale
factors $e^{\lambda_i} \;$, and $5 : 12345$ with scale factors
$e^{\lambda_i'}$ and $W : 3$ with scale factors
$e^{\lambda_i''}$ obtained by the U dualities $\uparrow_3 T_4
T_5 \downarrow_3 \;$ and $\uparrow_3 T_1 T_2 \downarrow_3 \;$ on
$2 : 12 \;$. The $\lambda_i \;$, $\lambda_i' \;$, and
$\lambda_i''$ obey the obvious symmetry relations \footnote{ The
obvious symmetry relations can also be obtained as consequences
of U dualities $\uparrow_i T_j S T_j \downarrow_i \;$ for
suitable $i$ and $j$ under which $i$ and $j$ are interchanged:
$\lambda_i' = \lambda_j \;$, $\lambda_j' = \lambda_i \;$,
$\lambda_l' = \lambda_l \;$, $l \ne i, j \;$.}
\begin{eqnarray}
2 & : & 
\lambda_1 = \lambda_2 \; \; , \; \; \; 
\lambda_3  = \cdots = \lambda_{10} \label{2} \\
5 & : & 
\lambda_1' = \cdots = \lambda_5' \; \; , \; \; \;
\lambda_6' = \cdots = \lambda_{10}' \label{5} \\
W & : & 
\lambda_3''  \; \; , \; \; \;
\lambda_1'' = \lambda_2'' = \lambda_4'' = \cdots 
= \lambda_{10}'' \label{W} \; \; .
\end{eqnarray}
The $\lambda_i'$ obtained from equations (\ref{ijk}), with $(i,
j, k) = (3, 4, 5)$ and hence $\lambda = \lambda_{10} \;$, are
\begin{eqnarray}
& & \lambda'_1 = \lambda'_2 = \lambda_1 + \lambda_{10}
\; \; , \; \; \; 
\lambda'_3 = \lambda'_4 = \lambda'_5 = - \lambda_{10} 
\nonumber \\
& & \lambda'_6 = \cdots = \lambda'_{10} = 2 \lambda_{10} 
\; \; . \label{1to1}
\end{eqnarray}
The obvious symmetry relation $\lambda'_1 = \lambda'_3 \;$, and
the above equations, imply further relations among $\lambda_i
\;$, and among $\lambda_i' \;$. Similarly for $\lambda_i''$
also. The extra relations for $\lambda_i \;$, $\lambda_i' \;$,
and $\lambda_i'' \;$, which thus follow from U duality, are
\begin{equation}\label{1x}
\lambda_1 + 2 \lambda_{10} = 0 \; \; , \; \; \; 
2 \lambda'_3 + \lambda'_{10} = 0 \; \; , \; \; \; 
\lambda_{10}'' = 0 \; \; .
\end{equation}

Consequences of U duality can be similarly obtained for
intersecting configurations also by this method: First consider
a pair of configurations, with scale factors $e^{\lambda_i}$ and
$e^{\lambda_i'} \;$, which are related by U duality; write down
the obvious symmetry relations for $\lambda_i$ and $\lambda_i'$
which may be obtained by inspection or by applying U dualities
$\uparrow_i T_j S T_j \downarrow_i \;$ with suitable $i, j \;$;
the U duality equations (\ref{ijk}) relate $\lambda_i'$ to
$\lambda_i$; these relations, and the obvious symmetry relations
for $\lambda_i' \;$, then imply further relations among
$\lambda_i' \;$, and among $\lambda_i \;$, as illustrated in the
example above. These extra relations are the consequences of U
duality. We will see shortly the sense in which all the
relations among $\lambda_i$ -- the obvious symmetry relations as
well as the U duality ones -- given in this paper may be taken
to be satisfied.

This method is simple and yet powerful. It yields all the
relations among $\lambda_i$ and shows, as may be expected, that
$\lambda_i$ for $N$ -- charge configurations are characterised
by $N$ independent functions. For example, consider the
configuration $W555 : (5, 12345, 23567, 14567)$ with scale
factors $e^{\lambda_i'}$ obtained by the U duality $\uparrow_5
T_1 T_2 \downarrow_5 \;$ on $2255 : (12, 34, 13567, 24567)$ with
scale factors $e^{\lambda_i} \;$. The obvious symmetry relations
for $\lambda_i$ and $\lambda_i'$ are
\begin{eqnarray}
2255 & : &  
\lambda_1 \; \; , \; \; \; \lambda_2 \; \; , \; \; \;
\lambda_3 \; \; , \; \; \; \lambda_4 \; \; , \; \; \;
\lambda_5 = \lambda_6 = \lambda_7 \; \; , \; \; \; 
\lambda_8 = \lambda_9 = \lambda_{10} \label{2255} \\ 
W555 & : & 
\lambda_1' = \lambda_4' \; \; , \; \; \;
\lambda_2' = \lambda_3' \; \; , \; \; \;
\lambda_5' \; \; , \; \; \; 
\lambda_6' = \lambda_7' \; \; , \; \; \;
\lambda_8' = \lambda_9' = \lambda_{10}' \; \; . \label{W555}
\end{eqnarray}
Expressing $\lambda_i'$ in terms of $\lambda_i$ using equations
(\ref{ijk}), and enforcing the obvious symmetry relations
$\lambda'_1 = \lambda'_4$ and $\lambda'_2 = \lambda'_3 \;$, 
then yields the U duality relations
\begin{equation}\label{4x}
\lambda_1 + \lambda_4 + \lambda_5 =
\lambda_2 + \lambda_3 + \lambda_5 = 0 \; \; , \; \; \; 
\lambda'_1 + \lambda'_2 + \lambda'_6 = 0 
\end{equation}
which also show that $\lambda_i \;$, and similarly $\lambda_i'
\;$, can be characterised by four independent functions. Similar
relations for other configurations can also be obtained
straightforwardly by this method, but will not be presented here
since they can all be obtained from equation (\ref{p1}) and an
ansatz for energy momentum tensor, given below.

\vspace{4ex}

{\bf 3. Consequences of U dualities for energy momentum tensor}

\vspace{2ex}

The energy momentum tensor for the $N$ -- charge intersecting
brane antibrane configurations, which are smeared uniformly in
the common transverse space, is of the form $T^\mu \; _\nu =
diag \; (- \rho, \; p_i) \;$. Then, for the metric given by
equation (\ref{ds}), Einstein's equations $R_{\mu \nu} -
\frac{1} {2} g_{\mu \nu} R = T_{\mu \nu}$ in natural units with
$8 \pi G = 1$ become
\begin{equation}\label{eom}
\dot{\Lambda}^2 - \sum_i \dot{\lambda}_i^2 = 2 \rho 
\; \; \; , \; \; \; \;
\ddot{\lambda}_i + \dot{\lambda}_i \dot{\Lambda}
- \ddot{\Lambda} - \dot{\Lambda}^2 = - \rho + p_i
\; \; \; , \; \; \; \Lambda = \sum_i \lambda_i 
\end{equation}
where $i = 1, 2, \cdots, 10$ and overdots denote derivatives
with respect to the physical time $t \;$. It follows from the
above equations that
\begin{equation}\label{fi}
\ddot{\lambda}_i + \dot{\lambda}_i \dot{\Lambda}
= p_i + \frac{\rho - P}{9} \equiv f_i 
\; \; \; , \; \; \; P = \sum_i p_i \; \; .
\end{equation} 
Note that if $\lambda_i$ satisfy a relation $\sum_i k_i
\lambda_i = 0 \;$, where $k_i$ are constants, then $\rho$ and
$p_i$ must be such that the functions $f_i$ defined above
satisfy $\sum_i k_i f_i = 0 \;$. 

Conversely, if $\rho$ and $p_i$ are such that $\sum_i k_i f_i =
0 \;$ then $\sum_i k_i \left( \ddot{\lambda}_i + \dot{\lambda}_i
\dot{\Lambda} \right) = 0 \;$. It then follows that $\sum_i k_i
\dot{\lambda}_i = K e^{- \Lambda}$ where $K$ is an integration
constant. {\bf (i)} If $e^\Lambda$ grows sufficiently fast in
the limit $t \to \infty \;$, namely if $e^\Lambda \simeq
t^\alpha$ with $\alpha > 1 \;$, then we have $\sum_i k_i
\lambda_i \simeq constant$ in this limit, independently of the
initial conditions. If $\alpha \le 1$ \footnote{For Kasner type
vacuum solutions, $\alpha = 1$.} then $\sum_i k_i \lambda_i$ is
a function of $t$ in the limit $t \to \infty$ also, and depends
on initial conditions. {\bf (ii)} In such cases, we impose the
condition $\sum_i k_i \dot{\lambda}_i = 0$ at some initial
time. Then $K = 0 \;$, and we have $\sum_i k_i \lambda_i =
constant$ for all time $t \;$. With no loss of generality, this
$constant$ can be set to zero by coordinate rescaling. It is in
the sense {\bf (i)} or {\bf (ii)} that $\sum_i k_i f_i = 0$
implies $\sum_i k_i \lambda_i = 0 \;$.

Thus, all the relations among $\lambda_i$ -- the obvious
symmetry relations as well as the U duality ones -- given in
this paper may be taken to be satisfied in the sense {\bf (i)}
or {\bf (ii)}, explained above. Which one applies to which
configuration can be decided by assuming the relations $\sum_i
k_i f_i = 0 \;$, and finding $\alpha$ from the solutions in the
limit $t \to \infty \;$. \footnote{ Obtaining the solutions in
the general case is difficult. For a class of N charge solutions
given in section {\bf 5}, see also \cite{cs, k3}, it turns out
that $\alpha \ge 1$ always, and $\alpha > 1$ for $N > 1$ and
$\alpha = 1$ for $N = 1 \;$. If we assume that this is the
generic behaviour in general then {\bf (i)} applies for $N > 1$
cases, and {\bf (ii)} for $N = 1$ cases.}

Consider the configuration $2: 12 \;$. It is natural to assume
that $p_1 = p_2 \equiv p_\parallel$ and $p_3 = \cdots = p_{10}
\equiv p_\perp \;$. Then, $f_i$ obey the relations $f_1 = f_2$
and $f_3 = \cdots = f_{10}$. Further, let $f_i$ obey the
relation $f_1 + 2 f_{10} = 0 \;$ also so that $\lambda_i$ may
satisfy, in the sense explained above, the obvious symmetry
relations and the U duality one given in equations (\ref{2}) and
(\ref{1x}). Using the definition of $f_i \;$, it follows easily
that $\rho + p_\parallel = 2 p_\perp \;$. A similar analysis for
5 branes and waves then shows that the U duality relations in
(\ref{1x}) imply that that the corresponding $\rho$ and $p_i$
obey a relation which we write as
\begin{equation}\label{p1}
p_\parallel = z \; (\rho - p_\perp) + p_\perp 
\end{equation}
where $\parallel$ or $\perp$ denotes directions parallel or
transverse to branes/wave, and $z = - 1$ for 2 branes and 5
branes and $= + 1$ for waves.

Equation (\ref{p1}), which is obtained as a consequence of U
dualities in M theory, is one of the main results of this
paper. It determines $p_\parallel$ in terms of $\rho$ and
$p_\perp$ which are, in general, functions of the brane and
antibrane charges $q$ and $\bar{q} \;$. \footnote{ The brane
charge $q \propto n \tau V$ where $n$, $\tau$, and $V$ denote
the number, tension, and volume of the branes. Similarly for
antibranes. For waves, $\tau V$ is to be replaced by $\frac{1}
{R}$ where $R$ is the size of the wave direction. Also, note
that if the common transverse space is compact then the net
charges must vanish as follows from Gauss's law. Hence, we
implicitly assume the common transverse space to be non compact
in the general case where the net charges do not vanish.} These
functions $\rho(q, \bar{q})$ and $p_\perp(q, \bar{q})$ may be
determined, in principle, by brane antibrane dynamics. $\rho(q,
\bar{q})$ and $p_\perp(q, \bar{q})$ must be same for 2 branes, 5
branes, and waves as follows from U dualities; but $p_\parallel$
will be different and is given by equation (\ref{p1}). If $q =
\bar{q} \;$, {\em i.e.} if the net charge vanishes, then
$p_\perp$ and $p_\parallel$ may be thought of as functions of
$\rho \;$.

Consider now $N$ -- charge intersecting configurations with $N >
1 \;$. We assume that the corresponding energy momentum tensor
is given, just as in the black hole case \cite{addn}, by the
ansatz
\begin{equation}\label{ansatz}
T^\mu \; _\nu = \sum_{I = 1}^N  T^\mu \; _{\nu (I)}
\; \; \; \Longrightarrow \; \; \; 
\rho = \sum_{I = 1}^N  \rho_{(I)}\; \; , \; \; \; 
p_i = \sum_{I = 1}^N p_{i (I)} 
\end{equation}
where $T^\mu \; _{\nu (I)} = diag \; (- \rho_{(I)}, \; p_{i
(I)}) \;$ is the energy momentum tensor of the $I^{th}$ type of
branes/wave, and $p_{i (I)}$ are given in terms of $p_{\parallel
(I)}$ and $p_{\perp (I)}$ which obey the relation (\ref{p1})
with $z = z_{(I)} = - 1$ for branes and $= + 1$ for waves.
\footnote{ The relation (\ref{p1}) among $\rho_{(I)} \;$,
$p_{\parallel (I)} \;$, and $p_{\perp (I)}$ may also be taken
as part of the ansatz, so that the $N > 1$ case is completely
independent of the $N = 1$ case. Then, the relations among
$\lambda_i$ are likely to be satisfied in the sense {\bf (i)}
explained earlier, see footnote 6, namely in the limit $t
\to \infty$, and independently of the initial conditions.}

It is straightforward to show that the above ansatz for $(\rho,
p_i)$ implies the obvious symmetry relations and, since
$(\rho_{(I)}, \; p_{i (I)})$ obey equation (\ref{p1}) with $z =
z_{(I)} \;$, also the U duality relations among $\lambda_i
\;$. For example, $(\rho, p_i)$ obtained from equation
(\ref{ansatz}) for the configuration $2255 : (12, 34, 13567,
24567)$ are
\begin{eqnarray}
\rho & = & \rho_{(1)} + \rho_{(2)} + \rho_{(3)} + \rho_{(4)}
\nonumber \\
p_1 & = & p_{\parallel (1)} + p_{\perp (2)} + p_{\parallel (3)}
+ p_{\perp (4)} \nonumber \\
p_2 & = & p_{\parallel (1)} + p_{\perp (2)} + p_{\perp (3)} +
p_{\parallel (4)} \nonumber \\
p_3 & = & p_{\perp (1)} + p_{\parallel (2)} + p_{\parallel (3)}
+ p_{\perp (4)} \nonumber \\
p_4 & = & p_{\perp (1)} + p_{\parallel (2)} + p_{\perp (3)} +
p_{\parallel (4)} \nonumber \\
p_5 = p_6 = p_7 & = & p_{\perp (1)} + p_{\perp (2)} +
p_{\parallel (3)} + p_{\parallel (4)} \nonumber \\
p_8 = p_9 = p_{10} & = & p_{\perp (1)} + p_{\perp (2)} +
p_{\perp (3)} + p_{\perp (4)} \label{4rhop} 
\end{eqnarray}
where $p_{\parallel (I)}$ is given by equation (\ref{p1}) with
$z_{(I)} = - 1$ for $I = 1, 2, 3, 4 \;$. It follows after some
algebra that $f_i \;$, given by equation (\ref{fi}), obey the
relations
\begin{eqnarray}
& & f_5 = f_6 = f_7 \; \; \; , \; \; \; f_8 = f_9 = f_{10}
\nonumber \\
& & f_1 + f_4 + f_5 = f_2 + f_3 + f_5 = 0 \label{4fi} 
\end{eqnarray}
which in turn imply, in the sense explained below equation
(\ref{fi}), the obvious symmetry relations and the U duality
ones for $\lambda_i$ given in equations (\ref{2255}) and
(\ref{4x}). We have similarly verified for various intersecting
configurations that the relations among $\lambda_i$ implied by
equation (\ref{p1}) and the ansatz in equation (\ref{ansatz})
are the same as those obtained directly by U dualities.

\vspace{4ex}

{\bf 4. Configurations with identical charges}  

\vspace{2ex}

Let $10^{th}$ direction be transverse to all of the $N$ types of
branes. Then $p_{10} = \sum_I p_{\perp (I)}$ and $p_i - p_{10} =
\sum_I \left( p_{i (I)} - p_{\perp (I)} \right) \;$. Note that
if $i^{th}$ direction is transverse to $I^{th}$ type of brane
then $p_{i (I)} = p_{\perp (I)} \;$. Hence, the sum in the
expression for $p_i - p_{10}$ is only over the remaining I,
denoted as $I \ni i \;$, for which $p_{i (I)} = p_{\parallel
(I)} \;$. Using equation (\ref{p1}) for $p_{\parallel (I)} \;$,
it now follows that
\begin{equation}\label{pip10} 
p_i - p_{10} = \sum_{I \ni i} z_{(I)} \; \left( \rho_{(I)} 
- p_{\perp (I)} \right) \; \; .
\end{equation} 

Consider now configurations with identical brane and antibrane
charges, namely with $q_1 = \cdots = q_N$ and $\bar{q}_1 =
\cdots = \bar{q}_N \;$. Then, $\rho_{(1)} = \cdots = \rho_{(N)}$
and $p_{\perp (1)} = \cdots = p_{\perp (N)} \;$. Therefore,
$\frac{\rho_{(I)} - p_{\perp (I)}} {\rho - p_{10}} =
\frac{1}{N}$ and it follows from equation (\ref{pip10}) that the
ratios $\frac{p_i - p_{10}} {\rho - p_{10}}$ are constants, say
$z_i\;$, and are given by
\begin{equation}\label{zi} 
z_i = \frac{p_i - p_{10}} {\rho - p_{10}} = 
\sum_{I \ni i} \frac{z_{(I)}}{N} \; . 
\end{equation}
Since $z_{(I)} = - 1$ for branes and $= + 1$ for waves, we have
that, in N -- charge configurations with identical charges, each
type of branes wrapping $i^{th}$ direction contributes $- \;
\frac{1}{N}$ to $z_i \;$, whereas a wave in that direction if
present contributes $+ \; \frac{1}{N}$ to $z_i$; the net $z_i$
is then the sum of all these contributions. $z_i = 0$ if there
are no branes or wave along $i^{th}$ direction.

From the definition of $z_i \;$, it follows that $\rho - p_i =
(1 - z_i) (\rho - p_{10}) \;$. Using this in equation
(\ref{eom}) for $\lambda_i \;$, and after some algebra, it
follows that
\begin{eqnarray} 
\ddot{\Lambda} + \dot{\Lambda}^2 & = & \left(
\frac{10 - \sum_j z_j}{9} \right) \; (\rho - p_{10}) \\
\ddot{\lambda}_i + \dot{\lambda}_i \dot{\Lambda} & = & 
l_i \; \left( \ddot{\Lambda} + \dot{\Lambda}^2 \right)
\; \; \; , \; \; \; 
l_i = 1 - \frac{9 (1 - z_i)}{10 - \sum_j z_j} \; \; .
\label{li}
\end{eqnarray} 
Since $z_i$ and, hence, $l_i$ are constants, equation (\ref{li})
implies that $\lambda_i = l_i \; \Lambda$ in the sense explained
below equation (\ref{fi}), {\em i.e.} in the limit $e^\Lambda
\to \infty \;$, equivalently $t \to \infty \;$, and upto
coordinate rescaling. For a given configuration with identical
charges, the constants $l_i$ can be calculated using equations
(\ref{zi}) and (\ref{li}). It can be verified that $l_i$ thus
obtained are the same as those obtained by applying U dualities
directly to configurations with identical charges. Note here
that the obvious symmetry relations are enhanced when charges
are identical.

Note that the constants $(z_i, l_i)$ and the relation $\lambda_i
= l_i \; \Lambda$ depend only on brane and antibrane charges
being identical; in particular, they are independent of the
details of the functions $\rho(q, \bar{q})$ and $p_\perp(q,
\bar{q})$ and, hence, of the details of the brane antibrane
dynamics that determines them.

The most interesting case is the four charge configuration 2255
or W555 with identical charges, for which $z_1 = \cdots = z_7 =
- \frac{2}{4} \;$, $z_8 = z_9 = z_{10} = 0 \;$, $l_1 = \cdots =
l_7 = 0 \;$, and $l_8 = l_9 = l_{10} = \frac{1}{3} \;$. Hence,
the sizes of the brane directions $(x^1, \cdots, x^7)$ become
constant in the limit $t \to \infty \;$. It thus follows that
the configuration 2255 or W555 with identical charges leads
asymptotically to an effective $(3 + 1)$ -- dimensional
expanding universe. Note that this result follows as a
consequence of U dualities in M theory and, as explained above,
is independent of the details of the brane antibrane dynamics
that determines the functions $\rho(q, \bar{q})$ and $p_\perp(q,
\bar{q}) \;$.

Similarly, for the three charge configuration 222 or 2W5 with
identical charges, $z_1 = \cdots = z_6 = - \frac{1}{3} \;$, $z_7
= \cdots = z_{10} = 0 \;$, $l_1 = \cdots = l_6 = 0 \;$, and $l_8
= \cdots = l_{10} = \frac{1}{4}$ leading, as before, to an
effective $(4 + 1)$ -- dimensional expanding universe.

\vspace{4ex}

{\bf 5. Configurations with vanishing net charges} 

\vspace{2ex}

It is known that, for a given energy, the entropy of the
intersecting brane antibrane configurations, which are localised
in the common transverse space and describe black holes, is
maximum when the charges are identical and the net charges all
vanish, {\em i.e.} $q_1 = \cdots = q_N = \bar{q}_1 = \cdots =
\bar{q}_N$ \cite{cs, hm, ks}. We assume this to be the case also
for the intersecting configurations which are smeared uniformly
in the common transverse space. Hence, in the following, we
focus on such configurations since, for entropic reasons, they
are likely to dominate the universe.

If the net charges all vanish, {\em i.e.} if $q_I = \bar{q}_I$
for $I = 1, \cdots, N \;$, then $p_\perp$ may be thought of as a
function of $\rho \;$. We assume that $p_\perp = w \rho$ where
$w$ is a constant in the range $- 1 \le w \le 1 \;$. Then,
equation (\ref{p1}) becomes
\begin{equation}\label{p1w}
p_\parallel = \left( z \; (1 - w) + w \right) \; \rho \; \; .
\end{equation} 
The energy momentum tensor for the N -- charge configuration is
given by the ansatz in equation (\ref{ansatz}) and depends on
$w$ and $\rho_{(1)} , \cdots , \rho_{(N)} \;$. If the charges
are also identical then we have $q_1 = \cdots = q_N = \bar{q}_1
= \cdots = \bar{q}_N \;$. Equivalently $\rho_{(1)} = \cdots =
\rho_{(N)}$ and $p_{\perp (1)} = \cdots = p_{\perp (N)} \;$. The
pressure $p_i$ is then given by
\begin{equation}\label{wi}
p_i = w_i \rho \; \; , \; \; \; 
w_i = z_i \; (1 - w) + w 
\end{equation} 
where the constants $z_i$ are given by equation (\ref{zi}). The
energy momentum tensor now depends on $w$ and $\rho$ only.

General solutions to the equations of motion (\ref{eom}) in this
case can be obtained from those for anisotropic universe given
recently by Chowdhury and Mathur in \cite{cs}. In this context,
note that Chowdhury and Mathur also derive the energy momentum
tensor for brane antibrane configurations in a certain
approximation and obtain the corresponding $w_i \;$, which can
also be obtained from the above expressions by setting $w = 0
\;$.

Here, we present the solutions for a few N -- charge
configurations in the asymptotic limit $t \to \infty \;$. See
\cite{cs, k3} for details. The functions $f_i$ in equation
(\ref{fi}), see also equation (\ref{li}), are now given by
\begin{equation}\label{ci}
f_i = C_i \; \rho \; \; \; , \; \; \; 
C_i = w_i + \frac{1 - \sum_j w_j}{9} = 
\left( z_i + \frac{1 - \sum_j z_j}{9} \right) \; 
(1 - w) \; \; . 
\end{equation}
If $\sum_i (1 - w_i) C_i > 0 \;$ then the solutions in the limit
$t \to \infty$ are independent of initial conditions and are
given by
\begin{equation}\label{alphai}
e^{\lambda_i} \simeq t^{\alpha_i} \; \; \; , \; \; \;
\alpha_i = \frac{2 C_i}{\sum_j (1 + w_j) C_j} \; \; . 
\end{equation}

The exponents $\alpha_i$ thus describe the scale factors in the
asymptotic limit $t \to \infty \;$. These exponents $\alpha_i
\;$, and also $z_i$ which describe the equation of state
(\ref{wi}), can be obtained straightforwardly in terms of $w$
using equations (\ref{zi}), (\ref{ci}), and (\ref{alphai}) for
any N -- charge configurations for which charges are identical
and net charges vanish, {\em i.e.} for which $q_1 = \cdots = q_N
= \bar{q}_1 = \cdots = \bar{q}_N \;$, and for which $\sum_i (1 -
w_i) C_i > 0 \;$. \footnote{$\sum_i (1 - w_i) C_i = 0$ for $N =
1 \;$. Equation (\ref{alphai}) is still applicable if one
assumes, as seems physically reasonable, that this $0$ is
approached from above \cite{k3}.} In Table I we list $(- N z_i)$
and $\alpha_i$ for a few N -- charge intersecting
configurations. The coordinates are arranged so that $\alpha_1
\le \alpha_2 \le \cdots \le \alpha_{10} \;$.

\vspace{2ex}

\begin{tabular}{||c||c|c||} 
\hline \hline & & \\ 

& $- N z_i$
& $ \alpha_i$ \\ 
& & \\ \hline  \hline 

2255 & & \\ 
& $2, 2, 2, 2, 2, 2, 2, 0, 0, 0$ 
& $(0, 0, 0, 0, 0, 0, 0, 2, 2, 2) 
\; \frac{1}{3 (1 + w)}$ \\
555W & & \\ \hline \hline 

222 & & \\
& $1, 1, 1, 1, 1, 1, 0, 0, 0, 0$ 
& $(0, 0, 0, 0, 0, 0, 1, 1, 1, 1) 
\; \frac{1}{2 (1 + w)}$  \\
25W & & \\ \hline 

225 & & \\ 
& $2, 2, 1, 1, 1, 1, 1, 0, 0, 0$ 
& $(- 2, - 2, 1, 1, 1, 1, 1, 4, 4, 4) 
\; \frac{1}{7 + 6 w}$ \\
55W & & \\ \hline 

& & \\
255 & $2, 2, 2, 2, 2, 1, 1, 0, 0, 0$ 
& $(- 1, - 1, - 1, - 1, - 1, 2, 2, 5, 5, 5) 
\; \frac{1}{2 (4 + 3 w)}$ \\
& & \\ \hline 

& & \\
555 & 
$3, 2, 2, 2, 2, 2, 2, 0, 0, 0$ 
& $(- 1, 0, 0, 0, 0, 0, 0, 2, 2, 2) 
\; \frac{1}{3 + 2 w}$  \\
& & \\ \hline \hline 

& & \\
2W & $1, 0, 0, 0, 0, 0, 0, 0, 0, 0$ 
& $(- 2, 1, 1, 1, 1, 1, 1, 1, 1, 1) 
\; \frac{1}{4 + 3  w}$  \\
& & \\ \hline 

22 & & \\
& $1, 1, 1, 1, 0, 0, 0, 0, 0, 0$ 
& $(- 1, -1, -1, -1, 2, 2, 2, 2, 2, 2) 
\; \frac{1}{5 + 3 w}$ \\
5W & & \\ \hline 

& & \\ 
55 & $2, 2, 2, 1, 1, 1, 1, 0, 0, 0$ 
& $(- 2, - 2, - 2, 1, 1, 1, 1, 4, 4, 4) 
\; \frac{1}{7 + 3 w}$  \\
& & \\ \hline 

& & \\ 
25 & $2, 1, 1, 1, 1, 1, 0, 0, 0, 0$ 
& $(- 1, 0, 0, 0, 0, 0, 1, 1, 1, 1) 
\; \frac{1}{2 + w}$ \\
& & \\ \hline \hline 




\end{tabular}

\vspace{2ex}

{\bf Table I :} 
{\em $z_i$ and $\alpha_i \;$, calculated using equations
(\ref{zi}), (\ref{ci}), and (\ref{alphai}), for a few $N$ --
charge configurations with $q_1 = \cdots = q_N = \bar{q}_1 =
\cdots = \bar{q}_N \;$. The corresponding $w_i$ are given by
equation (\ref{wi}). The scale factors $e^{\lambda_i} \simeq
t^{\alpha_i}$ in the asymptotic limit $t \to \infty \;$.}

\vspace{2ex}

We now make a few remarks. 

{\bf (i)} 
For the configurations 2255 and W555 with $m = 3$ common
transverse directions, and for the $N = 3$ configurations 222
and 25W with $m = 4$ common transverse directions, the brane
directions which are taken to be toroidal remain constant in
size in the asymptotic limit $t \to \infty$ and the $m$ common
transverse directions expand. The scale factors for the
transverse directions are identical to those of a $(m + 1)$ --
dimensional expanding isotropic universe containing a perfect
fluid with the equation of state $p = w \rho \;$. This result
follows as a consequence of U dualities in M theory, which was
conjectured in \cite{k3} following different considerations.
Thus, in particular, the 4 -- charge configuration 2255 or W555
leads asymptotically to an effective $(3 + 1)$ -- dimensional
expanding isotropic universe with the six compact directions
remaining constant in size.

{\bf (ii)} 
Other three charge configurations in Table I can all be
transformed to the configurations 222 or 25W by repeated
applications of U dualities. The two charge configurations can
also be transformed similarly to the configuration $2W : (12,
2)$ which, in turn, can be transformed to string theory
configuration $FW :(2, 2) \;$, namely fundamental strings along
$x^2$ direction with waves, by dimensional reduction along $x^1$
direction. The corresponding string metric and the dilaton
$\phi$ can be obtained easily in the asymptotic limit $t \to
\infty$ and are given by
\begin{equation}\label{iia}
d s^2 \simeq - d \tilde{t}^2 + \sum_{i = 2}^{10} d x_i^2
\; \; \; , \; \; \; e^\phi \simeq 
\tilde{t}^{\; - \frac{1}{1 + w}}
\end{equation}
where $\tilde{t} \simeq t^{\frac{3 (1 + w)} {4 + 3 w}} \;$. In
the asymptotic limit, $\tilde{t} \to \infty \;$, the scale
factors in string metric become constant, and the effective
string coupling $e^\phi \to 0 \;$.

{\bf (iii)} 
Thus, for the above configurations, namely 2255, W555, 222, 25W
in M theory and FW in string theory, the scale factors and,
hence, the physical sizes of the compact directions remain
constant in the asymptotic limit $t \to \infty \;$. We assume
that these asymptotic constant sizes, which depend on initial
conditions, are suitably large so that higher order corrections
to the original action are negligible and, consequently, the
present results remain valid.

In contrast, for the remaining configurations in Table I, one or
more $\alpha_i < 0$ for the compact directions and, hence, the
corresponding sizes $\to 0 \;$ in the asymptotic limit $t \to
\infty \;$ independent of the initial conditions. New light
modes are then likely to appear in such cases, higher order
corrections to the original action are likely to be non
negligible \cite{k97, bfm} and, consequently, the present
results for such configurations are likely to be modified.

The correct description in such cases may be obtained, as
explained in detail in \cite{bfm}, by using S, T, U dualities
and transforming these configurations to the `safe' ones for
which $\alpha_i \ge 0 \;$, the equality now being allowed by our
assumption above about constant sizes. In \cite{bfm}, for Kasner
type solutions with no energy momentum tensor, such
transformations are shown to always exist. Here, for the
configurations given in Table I, we see that these
transformations exist also in the presence of energy momentum
tensor. These transformations can be easily obtained, the `safe'
configurations now being 2255, W555, 222, 25W in M theory and FW
in string theory.

{\bf (iv)} 
Among these `safe' configurations, the configuration 2255 or
W555 has maximum entropy for a given energy. See \cite{k3} for
details. Hence, the universe is likely to be dominated by such
configurations. They may, therefore, provide a detailed
realisation of the maximum entropic principle that we have
proposed recently in \cite{k2} to determine the number $(3 + 1)$
of large spacetime dimensions. Even otherwise, these
configurations provide, at the least, a model for our $(3 + 1)$
-- dimensional expanding universe.

\vspace{4ex}

{\bf 6. Conclusion} 

\vspace{2ex}

We now briefly summarise the results. We considered N -- charge
intersecting brane antibrane configurations smeared uniformly in
the common transverse space so that they may describe our
universe. The brane directions are taken to be toroidal. We
found that U dualities imply relations among the scale factors,
which then are characterised by N independent functions.

The energy momentum tensor for such a configuration is of the
form $T^\mu \; _\nu = diag \; (- \rho, \; p_i)$ which may be
determined, in principle, by brane antibrane dynamics. It
follows from Einstein's equations that, as a consequence of U
duality, $\rho$ and $p_i$ for $N = 1$ case obey a relation given
in equation (\ref{p1}). We then presented an ansatz for $T^\mu
\; _\nu$ for the N -- charge case which, as can be verified,
yields all the U duality relations among the scale factors.

We studied configurations with $q_1 = \cdots = q_N$ and
$\bar{q}_1 = \cdots = \bar{q}_N$ and found, among other things,
that the configuration 2255 or W555 leads asymptotically to an
effective $(3 + 1)$ -- dimensional expanding universe. This
result follows as a consequence of U dualities in M theory and
is independent of the details of the brane antibrane dynamics.

We studied configurations with $q_1 = \cdots = q_N = \bar{q}_1 =
\cdots = \bar{q}_N$ which, for entropic reasons, are likely to
dominate the universe. We assumed that $p_\perp = w \rho
\;$. General solutions to the equations of motion can then be
obtained from those given in \cite{cs}. We listed the asymptotic
solutions for a few N -- charge configurations. It is seen,
following the reasoning given in \cite{bfm}, that the `safe'
configurations are 2255, W555, 222, 25W in M theory and FW in
string theory. Among these, the configuration 2255 or W555 has
maximum entropy and, hence, is likely to dominate the universe
and may, therefore, provide a detailed realisation of the
maximum entropic principle proposed in \cite{k2}. Even
otherwise, these configurations provide, at the least, a model
for our $(3 + 1)$ -- dimensional expanding universe.

We conclude by mentioning a few issues for further study. 

It is important to obtain general solutions to the equations of
motion with no restriction on charges. This, however, requires
the knowledge of brane antibrane dynamics that determines
$\rho(q, \bar{q})$ and $p_\perp(q, \bar{q}) \;$. In the absence
of such a knowledge, one may perhaps proceed by making a
suitable ansatz for $\rho(q, \bar{q})$ and $p_\perp(q,
\bar{q}) \;$.

It may be of interest to obtain the gauge fields $A_{\mu \nu
\rho}$ corresponding to such a general ansatz or, atleast, for
the simpler ansatz $p_\perp = w \rho$ used in the present paper.

Here, we assumed the brane directions to be toroidal. It may
also be of interest to understand the consequences of U
dualities for more general topologies.

Configurations with $q_1 = \cdots = q_N = \bar{q}_1 = \cdots =
\bar{q}_N$ are likely to dominate the universe for entropic
reasons. Also, the reasonings given in \cite{bfm} are invoked
here in restricting the relevant configurations to the `safe'
ones. Given the importance of these configurations in
determining the number $(3 + 1)$ of large spacetime dimensions,
it is crucial to understand how such a condition on charges and
such a restriction to `safe' configurations emerge dynamically.


{\bf Acknowledgement:} 
We thank the refree for his/her comments clarifying a couple of
points.



\end{document}